\documentclass[runningheads]{llncs}
\usepackage{graphicx}
\usepackage{amsmath}
\begin{document}

\title{Software defect prediction with zero-inflated Poisson models\thanks{Supported by TIN2016-76956-C3-R QARE-BadgePeople-TESTEAMOS}}
%
%
\author{Daniel Rodriguez\inst{1}\orcidID{000-0002-2887-0185} \and
Javier Dolado\inst{2}\orcidID{0000-0002-3301-5650} \and
Javier Tuya\inst{3}\orcidID{0000-0002-1091-934X} \and
Dietmar Pfahl\inst{4}\orcidID{0000-0003-2400-501X}}
\authorrunning{D. Rodriguez et al.}
%
\institute{University of Alcala, \and
Dept of Computer Science, 28805 Alcal\'a de Henares, Madrid, Spain
\email{daniel.rodriguezg@uah.es}\\
\and
University of the Basque Country, Facultad de Inform\'atica, 20018 Donostia, Spain
\email{javier.dolado@ehu.es}\\
\and
University of Oviedo,  Campus of Viesques, 33204 Gij\'on, Asturias, Spain\\
\email{tuya@uniovi.es}
\and
University of Tartu, 50090 Tartu, Estonia\\
\email{dietmar.pfahl@ut.ee}}
\maketitle              
\begin{abstract}
In this work we apply several Poisson and zero-inflated models for software defect prediction. We apply different functions from several R packages such as \textit{pscl}, \textit{MASS}, \textit{R2Jags} and the recent \textit{glmmTMB}. We test the functions using the Equinox dataset. The results show that Zero-inflated models, fitted with either maximum likelihood estimation or with Bayesian approach, are slightly better than other models, using the AIC as selection criterion. 

\keywords{Software defect prediction  \and Zero-inflated models \and AIC.}
\end{abstract}
%
%
\section{Zero-inflated models for Software Defect Prediction}
Most software defects datasets follow a distribution with a large number of non-defective modules, i.e., modules with zero bugs. Therefore, these datasets are highly unbalanced. Furthermore, when there are defects, modules tend to have a low number of defects. Most works in the literature have addressed this problem as an unbalanced binary supervised classification problem, i.e., modules are either defective or non-defective no matter the number of defects (e.g.\cite{RodriguezHHDR2014}). In this work, we explore several Zero-inflated models (ZIP), including Bayesian estimation, to predict the number of defects in software defect datasets taking into account the previously stated imbalance problem. Although ZIP models have been explored in the past \cite{khoshgoftaar2005} here were compare different ZIP models with different criteria other than the p-value. 
\paragraph{Definition of the Zero-inflated Poisson model.} The zero-inflated model splits the governing equation for the dependent variable $Y$ in two processes as shown in Equation (1): the first one generates those extra zeros with probability {$\pi$}, and the second equation follows a Poisson distribution that generates the counts (some may also be zero)~\cite{zuur2016beginner}.

\begin{equation}\label{eq1}
\begin{split}
\Pr (y_j = 0) & = \pi + (1 - \pi) e^{-\lambda} \\ 
\Pr (y_j = h_i) & = (1 - \pi) \frac{\lambda^{h_i} e^{-\lambda}} {h_i!},\qquad h_i \ge 1
\end{split}
\end{equation}

\section{The Equinox dataset}
In this work, we use the Equinox framework described by D'Ambros et al~\cite{DAmb2010a}. This dataset is part of the \textit{Bug prediction dataset}\footnote{\texttt{http://bug.inf.usi.ch/}} and corresponds to a Java Framework included the Eclipse project. 

\begin{figure}
\centering
\includegraphics[width=.65\textwidth]{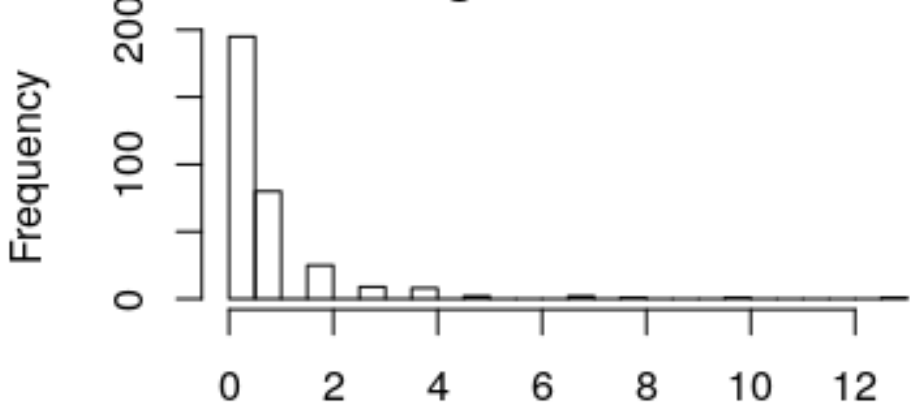}
\caption{The x-axis represents the number of bugs found in software modules. The y-axis represents the number of modules that contain x bugs.} 
\label{fig1}
\end{figure}

\begin{figure}
\centering
\includegraphics[width=.85\textwidth]{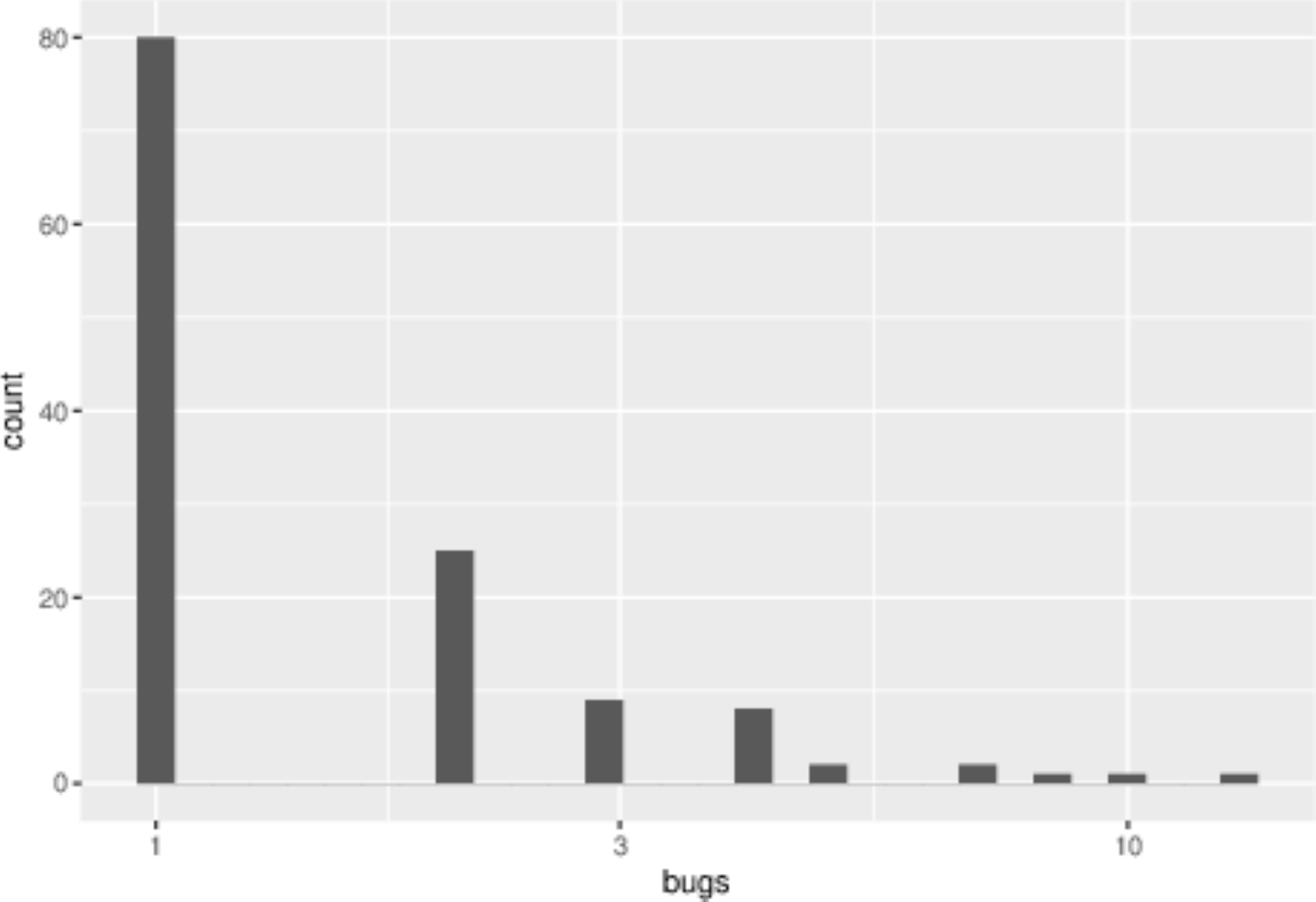}
\caption{After removing the zero bugs modules, the histogram still shows a big proportion of modules with few defects.} 
\label{fig2}
\end{figure}

Fig.~\ref{fig1} shows the high number of modules with no defects in the Equinox dataset. The second histogram in Fig.~\ref{fig2} shows the distribution of the non-zero values. 

\paragraph{Variable selection}
For the purpose of building the models, we performed several analyses using correlation, Akaike Information Criterion (AIC) and Bayesian Information Criterion (BIC),  giving as result that the best set of variables were: \textit{wmc}: weighted methods per class, that is simple the method count for a class; \textit{rfc}: response for a class is the set of methods that can potentially be executed in response to a message; \textit{cbo}: coupling between objects, i.e., number of classes to which a class is coupled; \textit{lcom}: lack of cohesion. For the excess count of zeros the variable, \textit{number of lines of code} (\textit{nloc}) was selected because it gives the user a clear understanding about what the source of zeros is. However, the variable \textit{nloc} can be safely replaced by \textit{wmc}, giving the latter slightly better results. 

\section{Simulation of ZIP models with R packages}
There are several R packages that can be used to analyze zero-inflated models such as \textit{pscl}, \textit{R2Jags}, \textit{MCMCglmm}, \textit{CARBayes}, \textit{R-INLA}, \textit{mgcv} and others. The last addition is the package \textit{glmmTMB} package~\cite{brooks2017glmmtmb,brooks2017modeling}. Some of them are based on the Maximum Likelihood Estimation and other packages use Bayesian simulation~\cite{mcelreath2018statistical}\cite{kruschke2014doing}.

\begin{table}
\centering
\caption{Summary of the results obtained with different R packages. }\label{tab1}
\begin{tabular}{l|l|l|l|l}
\hline
\textit{\textbf{Method}} & \textit{\textbf{AIC}} & \textit{\textbf{BIC}} & \textit{\textbf{R Package}} &  \textit{\textbf{\# Bugs predicted}}\\
\hline\hline
Regression & {\bfseries904.8354} & 927.5198 &MASS & 97.76806\\
Poisson  &  {\bfseries 632.1547}& 651.0584 & pscl & 188.7356\\
Poisson & {\bfseries 632.2} & 651.1 & glmmTMB & n.a \\
Poisson & {\bfseries 632.1547} & - & mgvc & -\\
Neg. binom.&  {\bfseries 644.5} & - & MASS & 195.8165\\
Neg. binom. & {\bfseries 628.6} & 651.2 & glmmTMB & n.a.\\ 
Neg. binom. & {\bfseries 628.5507} & - & mgvc & - \\
ZIP & {\bfseries 606.9155}& 633.3807 & pscl & 195.7924\\
ZIP & {\bfseries 606.9}& 633.4 & glmmTMB & n.a. \\
ZIP & {\bfseries 602.9} wmc & 629 & glmmTMB & n.a. \\
ZIP & - & {\bfseries DIC=622.5} & Bayes RJAGS & - \\
ZIP & {\bfseries 653.4149} & - & mgvc & - \\
ZIP & {\bfseries 647.9201} wmc & - & mgvc  & - \\
ZINB & {\bfseries 607.5639} & 637.8098 & pscl & 198.2048 \\
\hline
\end{tabular}
\end{table}

\paragraph{AIC, BIC, DIC.}
The AIC  and the BIC, or Schwarz information criterion, are common measures for model selection. The Deviance Information Critierion (DIC) is used in Bayesian model selection and is a generalization of the AIC.

Table~\ref{tab1} shows that the AIC is low in most of the ZIP models: \textit{pscl} and \textit{glmmTMB} give the same result. The Bayes regression performed with \textit{R2jags} gives similar coefficients (not shown here) as those of the \textit{pscl} version. The column "bugs predicted" has been computed for the models that had functions readily available. The ZIP model in \textit{pscl} predicts the same number of bugs as the actual value of the Equinox dataset, which was 195. 

\section{Conclusions}
Although ZIP models have presented good results across all R packages, more research is needed to generalize the validity of the ZIP approach. It makes sense to assume that many modules will not contain bugs because they have few lines of code or because they have been heavily tested in the past. Here we used only a single dataset and future work will include more datasets.

%
%
%
\bibliographystyle{splncs04}
\bibliography{refsMadsese1.bib}
%




\end{document}